\documentclass[conference]{IEEEtran}
\IEEEoverridecommandlockouts
\usepackage{todonotes}
\usepackage[utf8]{inputenc}
\usepackage{cite}
\usepackage{amsmath,amssymb,amsfonts}
\usepackage{mathtools} 
\usepackage{graphicx}
\usepackage{textcomp}
\usepackage{xcolor}
\usepackage{epstopdf}
\usepackage{algorithm,algpseudocode}
\def\BibTeX{{\rm B\kern-.05em{\sc i\kern-.025em b}\kern-.08em T\kern-.1667em\lower.7ex\hbox{E}\kern-.125emX}}

\def\BibTeX{{\rm B\kern-.05em{\sc i\kern-.025em b}\kern-.08em
		T\kern-.1667em\lower.7ex\hbox{E}\kern-.125emX}}

\DeclareMathOperator{\EX}{\mathbb{E}}
\DeclareMathAlphabet      {\mathbfit}{OML}{cmm}{b}{it}

\newtheorem{thm}{Theorem}

\begin{document}

\title{End-to-End Rate Enhancement in C-RAN Using Multi-Pair Two-Way Computation}
\author{
\IEEEauthorblockN{Mahmoud A. Hasabelnaby and Anas Chaaban}
\IEEEauthorblockA{School of Engineering, University of British Columbia, Kelowna, V1V1V7 BC, Canada\\
\{mahmoud.hasabelnaby,anas.chaaban\}@ubc.ca
}}

\maketitle

\begin{abstract}
Cloud radio-access networks (C-RAN) have been proposed as an enabling technology for keeping up with the requirements of next-generation wireless networks. 
Most existing works on C-RAN consider the uplink or the downlink separately. However, designing the uplink and the downlink jointly may bring additional advantages, especially if message source-destination information is taken into account. In this paper, this idea is demonstrated by considering pairwise message exchange between users in a C-RAN. A multi-pair two-way transmission scheme is proposed which targets maximizing the end-to-end user data rates. 
The achievable rate of this scheme is derived, optimized, and evaluated numerically. Results reveal that significant end-to-end rate improvement can be achieved using the proposed scheme compared to existing schemes.
\end{abstract}


\section{Introduction}
The cloud radio-access network (C-RAN) architecture is one of the methods {that enhance communication capabilities towards meeting} the critical requirements of next-generation wireless networks. 
However,the performance of a C-RAN is constrained by the limited capacity of fronthaul links that connect RRHs with the baseband processing unit (BBU) pool. This puts a constraint on the amount of information that can be exchanged between RRHs and the BBU pool. Therefore, advanced signal processing and relaying techniques are required in both uplink and downlink in order to make efficient use of the fronthaul links \cite{7096298}. This has been the topic of many studies recently as discussed next.

In the uplink, multiple users transmit their codewords to the RRHs. Different relaying strategies can be used to relay information from RRHs to the BBU pool, such as decode-and-forward (DF) \cite{quek_peng_simeone_yu_2017}, compute-and-forward (CoF) \cite{6034734}, and compress-and-forward (CF) \cite{6342931}. 
CoF is better than the CF under small fronthaul capacity values. However, the average performance of CF is better at moderate and high fronthaul capacities \cite{quek_peng_simeone_yu_2017,8956034,7465790}. In the downlink, transmission strategies that enable cooperation between RRHs include data-sharing strategies \cite{6920005,5997324}, compression based strategies \cite{quek_peng_simeone_yu_2017,6588350}, reverse compute-and-forward (RCoF) \cite{6283033}, and reverse quantized-compute-and-forward (RQCoF) \cite{6522158}.
Since the exact characterization of the downlink C-RAN capacity is still an open problem, most {works} optimize the schemes using uplink-downlink duality to achieve downlink rates greater than or equal to the uplink rates \cite{7541570,8635883}.

{Note that all aforementioned works study either the uplink or the downlink, separately. This may incur performance loss, especially in scenarios where intra-cloud message exchange is desired. This paper demonstrates this idea via studying a C-RAN with intra-cloud pairwise communication. Such a scenario can occur in video conferencing or gaming applications for instance.} A multi-pair two-way transmission scheme is proposed to maximize the end-to-end achievable rate. Using a lattice-based compression strategy, RRHs compress their observations and forward them to the BBU pool, which in turn computes integer linear combinations of codeword-pairs. This reduces the required number of computation steps at the BBU pool, thereby reducing the number of rate constraints. The BBU pool compresses the linear combinations and forwards them to the RRHs, which decompress the signals and transmit them to the users. Finally, users decode their desired message using their own messages as side information. The achievable rate of the scheme is derived, optimized, and evaluated numerically, showing superior performance to existing schemes in the literature.  

In the sequel, the following notations will be used. Column vectors and matrices are donated by boldface lowercase and uppercase letters, such as $\mathbf x$ and $\mathbf X$, respectively. The transpose of matrix $\mathbf X$ is donated by $\mathbf X^\top$. All the logarithms are to the base 2, and  $\log^+ (x) = \max (0,\log(x))$.

\section{System Model}
We consider a C-RAN consisting of $M$ single-antenna user pairs ($K=2M$ users), $L$ single-antenna RRHs,\footnote{This work can be extended to deal with MIMO systems.} and a central processor (BBU pool). Each RRH $\ell \in \{1, \dots,L\}$, is connected to the BBU pool via a digital noiseless fronthaul link with a limited capacity $C_{\ell}$. User pairs wish to communicate with each other using the C-RAN architecture. In other words, users $k,k^{'}\in \{1,\dots,K\}$, $k\neq k^{'}$, {exchange} messages with each other. As a result, a pairing matrix $\mathbf W$ with dimensions of $M \times K$ is defined, where $w_{m,k} \in \{0,1\}$ is a user-pair association indicator, i.e., $w_{m,k}=1$ if user $k$ belongs to pair $m$, and $w_{m,k}=0$ otherwise. Note that $\sum_{m=1}^{M}w_{m,k}=1$ for all $k$. The message exchange among the user-pairs is completed in two stages over $n$ channel uses each, an uplink phase and a downlink phase.\footnote{We assume that the uplink and downlink occur over the same frequency band using a half-duplex fashion} 

In the uplink, user $k\in\{1,\ldots,K\}$ encodes its message $\mathbfit{g}_k$ with rate $R_k$ into a codeword $\boldsymbol{x}_k^{ul}\in\mathbb{R}^{n}$ and sends it to the RRHs. This signal is subject to a power constraint $P_k^{ul}$. Note that we consider a real-valued transmission model for the sake of simplicity, bearing in mind that a complex-valued model can be addressed using the real-valued vector representation. The received signal at the RRHs is given by
\begin{equation}
\mathbf{Y}^{ul} = \mathbf{H}^{ul} \mathbf{X}^{ul} + \mathbf{Z}^{ul}
\end{equation}
where $\mathbf{Y}^{ul} = [\mathbfit{y}^{ul}_1, \dots, \mathbfit{y}^{ul}_L]^\top$, $\mathbfit{y}^{ul}_{\ell} \in \mathbb{R}^{n}$ is the received signal at RRH $\ell$, $\mathbf{H}^{ul} \in \mathbb{R}^{L \times K}$ is the {uplink} channel gain matrix between all users and RRHs, $\mathbf{X}^{ul} = [\mathbfit{x}^{ul}_1, \dots, \mathbfit{x}^{ul}_K]^\top$, and $\mathbf{Z}^{ul} \in \mathbb{R}^{L \times n}$ is additive white Gaussian noise with {independent and identically distributed (i.i.d.)  components with zero mean and unit variance $\mathcal{N}(0,1)$.} We assume that channels are Rayleigh fading and remain fixed across the transmission of a complete codeword (block fading). RRH $\ell\in\{1,\ldots,L\}$ processes the signal $\mathbfit{y}^{ul}_{\ell}$ into a message $e_{\ell}^{ul}$, and sends this message to the BBU pool using the fronthaul link. 

The downlink is described as follows. First, the BBU processes the received messages from all RRHs, then constructs messages $e_{\ell}^{dl}$, $\ell\in\{1,\ldots,L\}$, and sends $e_{\ell}^{dl}$ to RRH $\ell$ in the downlink using the fronthaul links. The RRH then processes $e_{\ell}^{dl}$ to construct a transmit signal $\boldsymbol{x}_{\ell}^{dl}\in\mathbb{R}^n$ with power constraint $p_{\ell}^{dl}\leq P_{\ell}^{dl}$ for transmission to the users. The received signals at users can be written in a matrix form as
\begin{equation}
\mathbf{Y}^{dl} = \mathbf{H}^{dl} \mathbf{X}^{dl} + \mathbf{Z}^{dl},
\end{equation}
where $\mathbf{Y}^{dl} = [\mathbfit{y}^{dl}_1, \dots, \mathbfit{y}^{dl}_K]^\top$, $\mathbf{H}^{dl} \in \mathbb{R}^{K \times L}$ is the downlink channel gain matrix between all RRHs and users, $\mathbf{X}^{dl} = [\mathbfit{x}^{dl}_1, \dots, \mathbfit{x}^{dl}_L]^\top$, and $\mathbf{Z}^{dl} \in \mathbb{R}^{K \times n}$ is additive white Gaussian noise with i.i.d $\mathcal{N}(0,1)$ components. Note that we assume channel reciprocity, i.e., $\mathbf{H}^{dl}=\mathbf{H}^{ul\top}$. Finally, each user uses its received signal in combination with its own message to decode the message of the paired user.

The goal is to design an uplink/downlink transmission scheme which takes this user pairing into account, and to derive its achievable rate. This is discussed in the following sections which discuss the uplink phase and the downlink phase, respectively.

\section{Uplink Transmission}
\label{2}
 
\subsection{Encoding at the users}
Using nested lattice coding \cite{6034734}, the lattice chain {$\Lambda_{c} \subseteq \Lambda_{f,K} \subseteq \ldots \subseteq \Lambda_{f,1}$} is generated, {consisting of $n$-dimensional lattices}. The coarse lattice $\Lambda_{c}$ is chosen to be good for channel coding and quantization simultaneously, whereas the fine lattices $\Lambda_{f_k}, k \in \{1,2,\ldots,K\}$ are good for quantization only. User $k$ generates its nested lattice codebook as {$\mathcal{C}_k^{ul} = \{\Lambda_{f_k} \cap \upsilon_{\Lambda_{c}}\}$}, where $\upsilon_{\Lambda_{c}}$ is the fundamental Voronoi region of the coarse lattice. Using a one-to-one mapping, it maps its message $\mathbfit{g}_k$ into a lattice point {$\mathbfit{s}_k^{ul} \in \mathcal{C}_k^{ul}$}. To make the transmitted signal independent on the lattice point, a random dither $\mathbfit{t}_k^{ul}$, uniformly distributed over $\upsilon_{\Lambda_{c}}$ and known to all nodes in the network, is added. The result is then reduced using a modulo-lattice operation with respect to $\Lambda_{c}$. {This leads to the signal} 
\begin{equation}
\mathbfit{u}_k^{ul} = (\mathbfit{s}_k^{ul} + \mathbfit{t}_k^{ul})~ \rm{mod}~ \Lambda_{\mathit{c}}
\end{equation}
which is then {scaled} by $b^{ul}_k$, and transmitted (i.e., $\mathbfit{x}_k^{ul} = b^{ul}_k \mathbfit{u}_k^{ul}$) to the RRHs. The transmitted signals from all users can be written as $\mathbf{X}^{ul} = \mathbf{B}^{ul} \mathbf{U}^{ul}$, where $\mathbf{B}^{ul} = \rm{diag}\mathit{(b_1^{ul}, \dots, b_K^{ul})}$ is a $K \times K$ scaling matrix and $\mathbf{U}^{ul} =[\mathbfit{u}_1^{ul}, \dots \mathbfit{u}_k^{ul}]^\top$ is the dithered codewords {matrix}. The uplink power constraint per user can be met by selecting the coarse lattice $\Lambda_{c}$ with second moment $\sigma^2 (\Lambda_{c}) = p^{ul}$ and {assigning proper scaling value} $b^{ul}_k$, {so that}  $\frac{1}{n}\EX[\|\mathbfit{x}_k^{ul}\|^2]= (b^{ul}_k)^2 p^{ul} \leq P^{ul}_k$.

\subsection{Compression at the RRHs}
The received signal $\boldsymbol{y}_{\ell}^{ul}$ at RRH $\ell$ is processed as follows. Given a lattice chain {$\tilde{\Lambda}_{c,1} \subseteq\ldots \subseteq \tilde{\Lambda}_{c,L} \subseteq \tilde{\Lambda}_{f,L} \subseteq \ldots \subseteq \tilde{\Lambda}_{f,1}$}, RRH $\ell$ generates its codebook as $\kappa_\ell^{ul} = \{\tilde{\Lambda}_{f,\ell} \cap \upsilon_{\tilde{\Lambda}_{c,\ell}}\}$ with rate equal to its fronthaul capacity $C_\ell$. The coarse lattice $\tilde{\Lambda}_{c,\ell}$ is good for channel coding and quantization simultaneously, thereby the probability of error can be neglected. In addition, the fine lattice $\tilde{\Lambda}_{f,\ell}$ must be good for quantization in order to be able to find a precise relationship between the quantization rates and distortion levels \cite{7745894}. Then, the $\ell$th RRH adds a random dither vector $\tilde{\mathbfit{t}}_{\ell}^{ul}$, uniformly distributed over $\upsilon_{\tilde{\Lambda}_{f,\ell}}$, to its observation to make the quantization error independent {of} the received signal $\mathbfit{y}_\ell^{ul}$. Moreover, using its generated codebook $\kappa_\ell^{ul}$, the $\ell$th RRH compresses its dithered observed signal using a lattice-based vector quantization as follows
\begin{equation}
\mathbfit{\ddot y}_\ell^{ul} = [Q_{\tilde{\Lambda}_{f,\ell}}(\mathbfit{y}^{ul}_{\ell}+\tilde{\mathbfit{t}}_\ell^{ul})]~\rm{mod}~ \tilde{\Lambda}_\mathit{c,\ell}.
\end{equation}
Then, RRH $\ell$ maps $\mathbfit{\ddot y}_\ell^{ul}$ to an index $e_\ell^{ul} \in \{1, \ldots, 2^{n C_\ell}\}$, and forwards it to the BBU pool via its fronthaul link.

\subsection{Decompression at the BBU pool}
Once the BBU pool receives the indices $e_1^{ul}, \ldots, e_L^{ul}$, it recovers $\mathbfit{\ddot y}_1^{ul}, \ldots, \mathbfit{\ddot y}_L^{ul}$, then subtracts the dithers, and reduces the result using the modulo-lattice operation with respect to $\tilde{\Lambda}_{c,\ell}$ as
\begin{equation}
\begin{split}
\mathbfit{\tilde{y}}_\ell^{ul} &= [\mathbfit{\ddot{y}}_\ell^{ul} - \tilde{\mathbfit{t}}_\ell^{ul}]~\rm{mod}~ \tilde{\Lambda}_{\mathit{c,\ell}} \\ 
&\stackrel{(a)}{=} [\mathbfit{y}_\ell^{ul} + \mathbfit{q}_\ell^{ul}]~\rm{mod}~ \tilde{\Lambda}_{\mathit{c,\ell}}  = [\mathbfit{\hat{y}}_{\mathit{\ell}}^{\mathit{ul}}]~\rm{mod}~ \tilde{\Lambda}_{\mathit{c,\ell}} 
\end{split}
\end{equation} 
where $(a)$ is obtained from the distributive law of the modulo-lattice operation, $\mathbfit{q}_\ell^{ul}=-[\mathbfit{y}_\ell^{ul} + \tilde{\mathbfit{t}}_\ell^{ul}]~\rm{mod}~ \tilde{\Lambda}_{\mathit{f,\ell}}$ is the compression distortion which is independent of $\mathbfit{y}_\ell^{ul}$ and uniformly distributed over $\upsilon_{\tilde{\Lambda}_{f,\ell}}$, 
and $\mathbfit{\hat{y}}_\ell^{ul} = \mathbfit{y}_\ell^{ul} + \mathbfit{q}_\ell^{ul}$. After that, the BBU pool proceeds to decode $L$ integer linear combinations {$\mathbfit{v}_{r,1}^{ul}, \ldots,\mathbfit{v}_{r,L}^{ul}$ as in \cite{8956034}, by exploiting the correlation between the received signals at all RRHs, where}
\begin{equation}
\label{vri_ul}
\mathbfit{v}_{r,i}^{ul}=\left[\sum\limits_{\ell=1}^{L} {a_{r,i,\ell}^{ul}} \mathbfit{\hat{y}}_\ell^{ul}\right]~\rm{mod}~ \tilde{\Lambda}_{\mathit{c,i}} ~
\mathit{\stackrel{(b)}{=}
\sum\limits_{\ell=1}^{L} {a_{r,i,\ell}^{ul}} \mathbfit{\hat{y}}_\ell^{ul}}
\end{equation}
$i \in\{1, \ldots, L\}$, $a_{r,i,\ell}^{ul} \in \mathbb{Z}$ is an integer coefficient, and $(b)$ is obtained with high probability (w.h.p.) if $\tilde{\Lambda}_{\mathit{c,i}}$ is good for channel coding and $\frac{1}{n}\EX[\|\mathbfit{v}_{r,i}^{ul}\|^2] < \sigma^2 (\tilde{\Lambda}_{c,i})$ \cite{7745894}. 
We write \eqref{vri_ul} in a matrix form as 
$\mathbf{V_{\mathit{r}}^{\mathit{ul}}} = \mathbf{A_{\mathit{r}}^{\mathit{ul}}} \mathbf{\hat{Y}^{\mathit{ul}}}$, where $\mathbf{A_{\mathit{r}}^{\mathit{ul}}}$ is a $L \times L$ full-rank integer coefficient matrix with full rank sub-matrices $\mathbf{A}_{r,[1:i]}^{ul}$  for $i \in\{1, \ldots, L\}$ and  $\mathbf{\hat{Y}}^{\mathit{ul}} = [\mathbfit{\hat{y}}^{ul}_1, \dots, \mathbfit{\hat{y}}^{ul}_L]^\top$. The integer coefficients can be selected to maximize the computation rate constraint, which allows us to increase the compression rate and decrease the
compression distortion. 

The compression rate at the $\ell$th RRH can be expressed as
\begin{align}
	\label{ul_compression}
	R_{r,\mathit{\ell}}^{ul}&=\\
	&\dfrac{1}{2}\log^+\left(\frac{\mathbfit{a}_{r,\ell}^{{ul}\top}(\mathbf{H}^{ul}\mathbf{B}^{ul}\mathbf{P}^{ul}\mathbf{B}^{{ul}^\top}\mathbf{H}^{{ul}^\top}+\mathbf{I}_L+\mathbf{D}^{ul})\mathbfit{a}_{r,\ell}^{ul}}{d_\ell^{ul}}\right)\nonumber
\end{align}
where $R_{r,\mathit{\ell}}^{ul} \leq C_\ell$, $\mathbfit{a}_{r,\ell}^{ul\top}$ is the $\ell$th row of $\mathbf{A_{\mathit{r}}^{\mathit{ul}}}$, $\mathbf{P}^{ul}=p^{ul}\mathbf{I}_K$ is a $K \times K$ diagonal matrix, $\mathbf{I}_L$ is a $L \times L$ identity matrix, $\mathbf{D}^{ul}$ is a $L \times L$ uplink compression distortion effective covariance matrix whose diagonal elements are equal to $\mathbfit{d}^{ul} ={[d_1^{ul},  \ldots, d_L^{ul}]}^\top$, and $d_\ell^{ul}$ is the distortion power level
of the fine lattice $\tilde{\Lambda}_{\mathit{f,\ell}}$ at RRH $\ell$. 

By multiplying $\mathbf{V_{\mathit{r}}^{\mathit{ul}}}$ by the inverse of the integer coefficient matrix $\mathbf{A_{\mathit{r, inv}}^{\mathit{ul}}}$, the BBU pool can recover 
\begin{equation}
\mathbf{\hat{Y}}^{ul}= \mathbf{H}^{ul} \mathbf{X}^{ul} + \mathbf{Z}^{ul}+\mathbf{Q}^{ul}
\end{equation}
where $\mathbf{Q}^{ul} = [\mathbfit{q}^{ul}_1, \dots, \mathbfit{q}^{ul}_L]^\top$.
\subsection{Multi-pair computation at the BBU pool}
The BBU pool proceeds to decode $M$ integer linear combinations of user-pairs' codewords (instead of decoding $K$ integer linear equations of users' individual codewords {as in {\cite{8956034}}}), to obtain 
\begin{equation}
\label{vpsij_ul}
\mathbfit{v}_{\psi,j}^{ul}=\left[\sum\limits_{m=1}^{M} {a_{\psi,j,m}^{ul}} \left[\sum\limits_{k=1}^{K} w_{m,k} \mathbfit{s}_k^{ul}\right]\right] ~\rm{mod}~\Lambda_{\mathit{c}},
\end{equation}
where $j \in\{1, \ldots, M\}$ and $a_{\psi,j,m}^{ul} \in \mathbb{Z}$ is an integer coefficient. Computing $M$ combinations instead of $K$ decreases the number {of} constraints on the computation rate, {which improves performance}. Note that \eqref{vpsij_ul} can be rewritten in matrix form as $\mathbf{V_{\mathit{\psi}}^{\mathit{ul}}} = [\mathbf{A_{\mathit{\psi}}^{\mathit{ul}}} \mathbf{W} \mathbf{S^{\mathit{ul}}}]~\rm{mod}~\Lambda_{\mathit{c}}$, where $\mathbf{A_{\mathit{\psi}}^{\mathit{ul}}}$ is an $M \times M$ integer coefficient matrix with a rank of $M=K/2$, and $\mathbf{S}^{ul} = [\mathbfit{s}^{ul}_1,  \dots, \mathbfit{s}^{ul}_K]^\top$. {This decoding can be done by linearly processing $\mathbf{\hat{Y}}^{ul}$ with a scaling equalizer $\boldsymbol{\rho}_j^{ul}$, removing the dither $\mathbfit{t}_{k}^{ul}$, and reducing the result modulo $\Lambda_{c}$ as} 
\begin{align}
\boldsymbol{\mu}_j^{ul}&=\left[\boldsymbol{\rho}_j^{ul\top}\mathbf{\hat{Y}}^{ul}-\sum\limits_{k=1}^{K}\mathbfit{t}_k^{ul}\right]~\rm{mod}~\Lambda_{\mathit{c}}\nonumber\\
&=\left[\boldsymbol{\rho}_j^{ul\top} \mathbf{H}^{ul}\mathbf{B}^{ul}\mathbf{S}^{ul} + \boldsymbol{\rho}_j^{ul\top}(\mathbf{Z}^{ul}+\mathbf{Q}^{ul})\right]~\rm{mod}~\Lambda_{\mathit{c}}\nonumber \\
&=[\underbrace{\mathbfit{a}_{\psi,j}^{ul\top}\mathbf{W} \mathbf{S}^{ul}}_{\text{desired signal}}+\mathbfit{z}_{\rm{eff},\mathit{j}}^{ul}]~\rm{mod}~\Lambda_{\mathit{c}}
\end{align}
from which the BBU pool computes $[\mathbfit{a}_{\psi,j}^{ul\top}\mathbf{W} \mathbf{S}^{ul}]~\rm{mod}~\Lambda_{\mathit{c}}$, where $\mathbfit{z}_{\rm{eff},\mathit{j}}^{ul}=(\boldsymbol{\rho}_j^{ul\top} \mathbf{H}^{ul}\mathbf{B}^{ul}- \mathbfit{a}_{\psi,j}^{ul\top}\mathbf{W})\mathbf{S}^{ul}+ \boldsymbol{\rho}_j^{ul\top}(\mathbf{Z}^{ul}+\mathbf{Q}^{ul})$ is the effective noise with power ${\sigma_j^{ul}}^2 = \frac{1}{n}\EX[\|\mathbfit{z}_{\rm{eff},\mathit{j}}^{ul}\|^2$
\begin{align}
\label{sigma_j_ul}
{\sigma_j^{ul}}^2 
&= {\|(\boldsymbol{\rho}_j^{ul\top} \mathbf{H}^{ul}\mathbf{B}^{ul}- \mathbfit{a}_{\psi,j}^{ul\top}\mathbf{W})({\mathbf{P}^{ul}})^{{\tfrac{1}{2}}}\|}^{2}+ \\\nonumber&~~~~~\boldsymbol{\rho}_j^{ul\top}(\mathbf{I}_L+\mathbf{D}^{ul})\boldsymbol{\rho}_j^{ul}
\end{align}

In order to minimize the effective variance in {\eqref{sigma_j_ul}}, $\boldsymbol{\rho}_j^{ul}$ is chosen as the MMSE {scaling equalizer} given by
\begin{equation}
\boldsymbol{\rho}_j^{ul\top} = \mathbfit{a}_{\psi,j}^{ul\top}\mathbf{W}\mathbf{P}^{ul}\mathbf{H^{\mathit{ul\top}}}{(\mathbf{H^{\mathit{ul}}} {\mathbf{P}^{ul}\mathbf{H^{\mathit{ul\top}}}}+\mathbf{I}_L+\mathbf{D}^{ul})}^{-1}
\end{equation} 

By substituting the MMSE solution into \eqref{sigma_j_ul} and applying the matrix inversion lemma, the effective noise power ${\sigma_j^{ul}}^2$ can be rewritten as
\begin{align}
	\label{sigma_j_ul2}
	{\sigma_j^{ul}}^2 = \mathbfit{a}_{\psi,j}^{ul\top}\mathbf{W}[\mathbf{F}^{ul}_\psi \mathbf{F}^{ul\top}_\psi]\mathbf{W^\top}\mathbfit{a}_{\psi,j}^{ul} = {\|\mathbf{F}^{ul}_\psi \mathbf{W^\top} \mathbfit{a}_{\psi,j}^{ul}\|}^2
\end{align}
where $\mathbf{F}^{ul}_\psi$ is the Cholesky decomposition satisfying $\mathbf{F}^{ul}_\psi \mathbf{F}^{ul\top}_\psi= ({\mathbf{P}^{ul}}^{-1}+\mathbf{H^{\mathit{ul\top}}}{(\mathbf{I}_L+\mathbf{D}^{ul})}^{-1}\mathbf{H^{\mathit{ul}}})^{-1}$.

Let user $k$ belong to pair $m_k$, i.e., $w_{m,k}=1$. Then, the achievable uplink computation rate for user $k$ can be expressed as 
\begin{equation}
\label{Rpsi_ul}
R^{ul}_{\mathit{k}} \leq \mathop{\hbox{min}}_{\substack{j\in\{1,\ldots,M\}\\{a}_{\psi,j,m_k}^{ul}\neq 0}}  \frac{1}{2} \log^+\left( \frac{p^{ul} }{{\sigma^{ul}_j}^2}\right),
\end{equation} 
where ${\sigma^{ul}_j}^2$ is given by \eqref{sigma_j_ul2}. Instead of recovering the original messages as in \cite{8956034}, the BBU compresses the previously computed equations directly and forwards them to the RRHs through the fronthaul links {as described next}. 

\section{Downlink Transmission}
\label{3}
The basic idea of the downlink is to employ a reverse-quantized-compute-and-forward scheme \cite{6522158}. 

\subsection{Compression at the BBU pool}
At first, the BBU pool uses the beamforming matrix $\mathbf{B}^{dl}$ with dimensions $L \times M$ to produce
\begin{equation}
\mathbf{S}^{dl} = \mathbf{B}^{dl}~ \mathbf{V}^{ul}_{\psi} = \mathbf{B}^{dl}[\mathbf{A}_{\psi}^{ul}\mathbf{W} \mathbf{S^{\mathit{ul}}}]~\rm{mod}~ \Lambda_{\mathit{c}} 
\end{equation} 
where $\mathbf{S}^{dl} = [\mathbfit{s}^{dl}_1, \dots, \mathbfit{s}^{dl}_L]^\top$. 
In order to enable each RRH to extract its desired quantized signal, the BBU pool pre-inverts the $\mathbf{S}^{dl}$ with $\mathbf{A}_{r,inv}^{dl}$ as follows
\begin{equation}
\mathbf{V}^{dl}_{r} = \mathbf{A}_{r,inv}^{dl} \mathbf{S^{\mathit{dl}}} 
\end{equation} 
where $\mathbf{V}^{dl}_{r} = [\mathbfit{v}^{dl}_1, \dots, \mathbfit{v}^{dl}_L]^\top$ and $\mathbf{A}_{r,inv}^{dl}$ is the inverse of the $L \times L$ full rank integer coefficient matrix $\mathbf{A}_{r}^{dl}$. Then, the BBU pool {uses a lattice chain $\hat{\Lambda}_{c,1}  \subseteq ... \subseteq \hat{\Lambda}_{c,L} \subseteq \hat{\Lambda}_{f}$,} where the coarse lattices and the fine lattice have the same properties as mentioned in the user encoding step. Next, the BBU pool adds a random dither matrix $\hat{\mathbf{T}}^{dl} = [\hat{\mathbfit{t}}^{dl}_1, \ldots, \hat{\mathbfit{t}}^{dl}_L]^\top$ to $\mathbf{V}^{dl}_{r}$ which is uniformly distributed over $\upsilon_{\hat{\Lambda}_{f}}$. The dithered output is then quantized as 
\begin{equation}
\mathbf{\hat{V}}^{dl}_r = Q_{\hat{\Lambda}_f}(\mathbf{V}^{dl}_{r}+\hat{\mathbf{T}}^{dl})
\end{equation}
where $Q_{\hat{\Lambda}_f}$ is applied to each row of the dithered matrix separately. The BBU pool proceeds to generate integer linear combinations $\mathbf{\tilde{V}}^{dl}_r = \mathbf{A}^{dl}_r \mathbf{\hat{V}}^{dl}_r$ and performs the modulo-lattice operation with respect to $\hat{\Lambda}_{c,i}$, $i \in \{1, \dots, L\}$ to each $i$th row in $\mathbf{\tilde{V}}^{dl}_r$ to obtain
\begin{equation}
\begin{split}
\mathbfit{\tilde{v}}^{dl}_{r,i} &= \left[\mathbfit{a}^{dl\top}_{r,i} \mathbf{\hat{V}}^{dl}_r\right]~\rm{mod}~\hat{\Lambda}_{\mathit{c,i}}\\
&=\left[\mathbfit{a}^{dl\top}_{r,i} Q_{\hat{\Lambda}_f}(\mathbf{V}^{dl}_{r}+\hat{\mathbf{T}}^{dl})\right]~\rm{mod}~\hat{\Lambda}_{\mathit{c,i}}
\end{split}
\end{equation}
Finally, the BBU pool maps its compressed linear equation $\mathbfit{\tilde{v}}^{dl}_{r,i}$ to an index $e_i^{dl} \in \{1,\ldots, 2^{n C_l}\}$, and forwards it to the $i$th RRH.

\subsection{Decompression at the RRHs}
Once the $\ell$th RRH receives the index $e_\ell^{dl}$, it recovers $\mathbfit{\tilde{v}}^{dl}_{r,\ell}$, then subtracts the dither $\hat{\mathbf{T}}^{dl}$, and reduces the result using the modulo-lattice operation with respect to $\hat{\Lambda}_{c,\ell}$ to obtain
\begin{align}
\mathbfit{x}_\ell^{dl} &= \left[\mathbfit{\tilde{v}}^{dl}_{r,\ell}-\mathbfit{a}^{dl\top}_{r,\ell} \hat{\mathbf{T}}^{dl}\right]~\rm{mod}~\hat{\Lambda}_{\mathit{c,\ell}}\nonumber\\
&=\left[\mathbfit{a}^{dl\top}_{r,\ell}(\mathbf{V}^{dl}_{r}+\hat{\mathbf{T}}^{dl}+\mathbf{Q}^{dl})-\mathbfit{a}^{dl\top}_{r,\ell} \hat{\mathbf{T}}^{dl}\right]~\rm{mod}~\hat{\Lambda}_{\mathit{c,\ell}}\nonumber\\
& \mathit{\stackrel{(c)}{=}} ~\mathbfit{s}^{dl\top}_{\ell}+\mathbfit{a}^{dl\top}_{r,\ell}\mathbf{Q}^{dl}
\end{align}
where $\mathbf{Q}^{dl}=[\mathbfit{q}^{dl}_1, \dots, \mathbfit{q}^{dl}_L]^\top$ is the downlink compression distortion with a $L \times L$ effective covariance matrix $\mathbf{D}^{dl}$ whose diagonal elements is equal to $\rm{diag}\mathit{(d_1^{dl}, \ldots, d_L^{dl})}$, and (c) is obtained w.h.p. if $\frac{1}{n}\EX[\|\mathbfit{x}_{\ell}^{dl}\|^2] < \sigma^2 (\hat{\Lambda}_{c,\ell})$. The downlink compression rate at RRH $\ell$ is given by
\begin{equation}
\label{comp_dl}
R_{r,\mathit{\ell}}^{dl}=\dfrac{1}{2}\log^+\left(\frac{\mathbfit{b}^{dl\top}_\ell\mathbf{P}^{ul}_\psi\mathbfit{b}^{dl}_\ell+ \mathbfit{a}_{r,\ell}^{dl\top} \mathbf{D}^{dl} \mathbfit{a}_{r,\ell}^{dl}}{d_\ell^{dl}}\right)
\end{equation} 
where $R_{r,\mathit{\ell}}^{dl} \leq C_\ell$ and $\mathbf{P}^{ul}_\psi=p^{ul}~\mathbf{I}_{M}$ is $M \times M$ diagonal power matrix whose diagonal elements are equal to $p^{ul}$. Finally, after the $\ell$th RRH recovers its desired signal, it broadcasts $\mathbfit{x}_\ell^{dl}\in \mathbb{R}^{n}$ to the users with power 
\begin{align}
\label{tilde_dl}
\hspace{-.2cm}\frac{1}{n}\EX[\|\mathbfit{x}_{\ell}^{dl}\|^2]=\mathbfit{b}^{dl\top}_\ell\mathbf{P}^{ul}_\psi\mathbfit{b}^{dl}_\ell+ \mathbfit{a}_{r,\ell}^{dl\top} \mathbf{D}^{dl} \mathbfit{a}_{r,\ell}^{dl} \triangleq {p}^{dl}_\ell\leq P^{dl}_\ell.
\end{align}

\subsection{Decoding at the users} 
The received signals at all the users can be written in a matrix form as
\begin{equation}
\mathbf{Y}^{dl} =\mathbf{H}^{dl}(\mathbf{B}^{dl}\mathbf{V}^{ul}_{\psi}+\mathbf{A}^{dl}_{r} \mathbf{Q}^{dl})+ \mathbf{Z}^{dl}.
\end{equation}
The $k$th user scales its received signal $\mathbfit{y}_k^{dl}$ by a linear {scaling} ${\rho}_k^{dl}$ {and reduces the result modulo $\Lambda_{c}$ as follows}
\begin{align}
\boldsymbol{\mu}_k^{dl}&=[{\rho}_k^{dl}\mathbfit{y}_k^{ul}]~{\rm{mod}}~\Lambda_{\mathit{c}}\nonumber\\
&=[{\rho}_k^{dl} \mathbfit{h}_k^{dl}(\mathbf{B}^{dl}\mathbf{V}^{ul}_{\psi}+\mathbf{A}^{dl}_{r}\mathbf{Q}^{dl}) + {\rho}_k^{dl} \mathbfit{z}_k^{dl}]~{\rm{mod}}~\Lambda_{\mathit{c}} \nonumber\\
&=[\underbrace{\mathbfit{a}_{\psi,k}^{dl}\mathbf{V}^{ul}_{\psi}}_{\text{intended signal}}+ \mathbfit{z}_{\rm{eff},\mathit{k}}^{dl}]~{\rm{mod}}~\Lambda_{\mathit{c}}
\end{align}
{where $\mathbfit{z}_{\rm{eff},\mathit{k}}^{dl}$ is the effective noise given by $({\rho}_k^{dl} \mathbfit{h}^{dl}_k\mathbf{B}^{dl}- \mathbfit{a}_{\psi,k}^{dl})\mathbf{V}^{ul}_{\psi}+ {\rho}_k^{dl} (\mathbfit{h}^{dl}_k \mathbf{A}^{dl}_{r}\mathbf{Q}^{dl}+ \mathbfit{z}^{dl}_k)$,} $\mathbfit{a}_{\psi,k}^{dl}$ is the $k$th row of $\mathbf{A}_{\psi}^{dl}$, a matrix with dimensions of $K \times M$ and rank of $M$, $\mathbf{A}_{\psi}^{dl} = \mathbf{W}^{\top} \mathbf{A}_{\psi,inv}^{ul}$, $\mathbf{A}_{\psi,inv}^{ul}$ is the inverse of $\mathbf{A}_{\psi}^{ul}$ matrix, $\mathbfit{h}^{dl}_k$ is the $k$th row of $\mathbf{H}^{dl}$, $\mathbfit{v}_{\psi,k}^{dl}=[\mathbfit{a}_{\psi,k}^{dl}\mathbf{V}^{ul}_{\psi}]~{\rm{mod}}~\Lambda_{\mathit{c}}$ is the $k$th user's intended signal that includes the sum of the {codewords of the user-pair $k,k^{'}\in \{1,\dots,K\}$, $k\neq k^{'}$. The power of the effective noise $\mathbfit{z}_{\rm{eff},\mathit{k}}^{dl}$ is given by}
\begin{align}
{\sigma_k^{dl}}^2
&=\frac{1}{n}\EX[\|\mathbfit{z}_{\rm{eff},\mathit{k}}^{dl}\|^2]\\ \nonumber 
&= {\|({\rho}_k^{dl} \mathbfit{h}^{dl}_k\mathbf{B}^{dl}- \mathbfit{a}_{\psi,k}^{dl})({\mathbf{P}^{ul}_\psi})^{{\tfrac{1}{2}}}\|}^{2}\\\nonumber
&\qquad\qquad+{{\rho}_k^{dl}}^{2}(\mathbfit{h}^{dl}_k\mathbf{A}^{dl}_r\mathbf{D}^{dl}\mathbf{A}^{dl^\top}_r\mathbfit{h}^{dl^\top}_k+1)
\end{align} 
This effective variance can be minimized by obtaining the MMSE coefficient for the linear {scaling equalizer} ${\rho}_k^{dl}$  as
\begin{equation}
\label{rhok_dl}
{\rho}_k^{dl} = \dfrac{\mathbfit{a}_{\psi,k}^{dl}\mathbf{P}^{ul}_\psi\mathbf{B}^{dl\top}\mathbfit{h}^{dl\top}_k}{\mathbfit{h}^{dl}_k(\mathbf{A}^{dl}_r\mathbf{D}^{dl}\mathbf{A}^{dl\top}_r+\mathbf{B}^{dl}\mathbf{P}^{ul}_\psi\mathbf{B}^{dl\top})\mathbfit{h}^{dl\top}_k+1}  
\end{equation} 

Finally, {user $k$ decodes $[\mathbfit{a}_{\psi,k}^{dl}\mathbf{V}^{ul}_{\psi}]~{\rm{mod}}~\Lambda_{\mathit{c}}$, and uses} its own codeword $[\mathbfit{s}_k^{ul}]~{\rm{mod}}~\Lambda_{\mathit{c}}$ as side information to recover its desired codeword $[\mathbfit{s}_{\mathit{k}'}^{ul}]~{\rm{mod}}~\Lambda_{\mathit{c}}$ 
as follows
\begin{align}
\mathbfit{s}_{desired,k}&=\left[[\mathbfit{a}_{\psi,k}^{dl}\mathbf{V}^{ul}_{\psi}]~{\rm{mod}}~\Lambda_{\mathit{c}}- \mathbfit{s}_k^{ul}\right]~{\rm{mod}}~\Lambda_{\mathit{c}}\nonumber\\
&=[\mathbfit{a}_{\psi,k}^{dl}\mathbf{A}_{\psi}^{ul}\mathbf{W} \mathbf{S}^{ul}- \mathbfit{s}_k^{ul}]~{\rm{mod}}~~\Lambda_{\mathit{c}}\nonumber\\
&= \left[\sum_{m=1}^{M}w_{m,k}(\sum\limits_{u=1}^{K}w_{m,u}\mathbfit{s}_u^{ul})- \mathbfit{s}_k^{ul}\right]~{\rm{mod}}~~\Lambda_{\mathit{c}}\nonumber\\
&=[\mathbfit{s}_k^{ul}+\mathbfit{s}_{k^{'}}^{ul}-\mathbfit{s}_k^{ul}]~{\rm{mod}}~~\Lambda_{\mathit{c}}\nonumber\\ 
&= [\mathbfit{s}_{k^{'}}^{ul}]~{\rm{mod}}~~\Lambda_{\mathit{c}}
\end{align}

Using this procedure, user $k$ downlink rate is given by
\begin{equation}
\label{Rpsi_dl}
R^{dl}_{\psi,\mathit{k}} = \frac{1}{2} \log^+( p^{ul} ({\sigma^{dl}_k}^2)^{-1}))
\end{equation}

At this point, 
we can summarize the end-to-end achievable rate of the proposed scheme as given next.

\begin{thm}
The end-to-end data rate of user $k$ achieved by the proposed scheme is given by
\begin{equation}
\label{EERate}
R_k = \min\{R^{ul}_{\psi,\mathit{k}},R^{dl}_{\psi,\mathit{k}}\},
\end{equation}
where $R^{ul}_{\psi,\mathit{k}}$ and $R^{dl}_{\psi,\mathit{k}}$ are given in \eqref{Rpsi_ul} and \eqref{Rpsi_dl}, respectively. 
\end{thm}
{\it Proof:}
This statement follows since the achievable end-to-end rate is bound by the smallest between the uplink rate and the downlink rate.

\section{End-to-End User-rate Optimization}
\label{4}
In this section, we propose an iterative multi-pair two-way rate optimization (MPTWR) algorithm to optimize the  end-to-end rate in \eqref{EERate}. The algorithm is carried in two steps, where the uplink and downlink user-rates are optimized iteratively. 

\subsection{Uplink Rate Optimization}
Given $\mathbf{H}^{ul}$, $\mathbf{P}^{ul}$ and $P_k^{ul}, k \in \{1, \dots, K\}$, the achievable uplink rate $\mathit{R}^{ul}_{\psi,\mathit{k}}$ can be optimized by assigning appropriate scaling matrix $\mathbf{B}^{ul}$, selecting proper full rank integer coefficient matrices $\mathbf{A}^{ul}_r$ and $\mathbf{A}^{ul}_\psi$, and selecting the uplink compression distortion covariance matrix $\mathbf{D}^{ul}$ to satisfy the fronthaul capacity constraint. The uplink optimization problem can be formulated as follows
\begin{align}
\mathop{\hbox{max}}_{\mathbf{A}^{ul}_r,{\mathbf{D}^{ul}},\mathbf{A}^{ul}_\psi}  & \sum\limits_{k=1}^{K}\mathit{R}^{ul}_{\psi,\mathit{k}} \cr {\hbox{subject to}}\quad &  (b^{ul}_k)^2 p^{ul} \leq P^{ul}_k ~\forall~ k \in \{1, 2, \dots, K\} \cr & \rm{rank}(\mathbf{A}^{\mathit{ul}}_\mathit{r})=\mathit{L},~~~~ \rm{rank}(\mathbf{A}^{\mathit{ul}}_\psi)=\mathit{M} \cr & 
\label{Constraint2}
\mathit{R}^{ul}_{r,\ell}\leq C_\ell ~\forall~ \ell \in \{1, 2, \dots, L\}
\end{align}
This optimization problem is a mixed-integer non-linear programming (MINLP) which is NP-hard problem. Therefore, we can solve (\ref{Constraint2}) by decoupling it into two separate steps and iterating over them. At first, for a fixed scaling matrix $\mathbf{B}^{ul}$, we start by selecting the proper integer coefficient matrices $\mathbf{A}^{ul}_r$ and $\mathbf{A}^{ul}_\psi$. This process is related to the Shortest Independent Vector Problem (SIVP) which is NP-hard. However, sub-optimal solutions can be obtained using the LLL algorithm \cite{Lenstra}. For the sake of simplicity, it is assumed that all RRHs choose equal distortion levels, i.e., $d_\ell^{ul} = d^{ul}$, $\ell \in \{1, 2, \dots, L\}$. To maximize the uplink rates, we start by initializing $d^{ul}$ to two extreme values and then calculate the corresponding $\mathbf{A}^{ul}_r$ using the LLL algorithm on $\mathbf{F}^{ul}_r$ which is the Cholesky decomposition satisfying $\mathbf{F}^{ul}_r \mathbf{F}^{ul\top}_r = \frac{1}{d^{ul}} \mathbf{H}^{ul}\mathbf{B}^{ul}\mathbf{P}^{ul}\mathbf{B}^{ul^\top}\mathbf{H}^{{ul}^\top}+\mathbf{I}_L(\frac{1}{d^{ul}}+1)$. Next, we update $d^{ul}$ using bisection until \eqref{Constraint2} is satisfied with equality. Finally, we use the obtained $d^{ul}$ to calculate $\mathbf{A}^{ul}_\psi$ using the LLL algorithm on $\mathbf{F}^{ul}_\psi$ defined after \eqref{sigma_j_ul2}. This is explained in detail in Algorithm 1. 

Second, under the assignment of $\mathbf{D}^{ul}$, $\mathbf{A}^{ul}_r$,  and $\mathbf{A}^{ul}_\psi$ matrices, the algorithm solves to find the appropriate precoding matrix $\mathbf{B}^{ul}$ using the barrier method which is a standard method to solve optimization problems with inequality constraints \cite{boyd_vandenberghe_2004} as shown in Algorithm 2. Finally, the algorithm iterate over the two sub-steps until convergence occurs. The results of this algorithm will be used as inputs to optimize the achievable downlink user-rate.
\begin{algorithm}[t]
	\caption{Iterative uplink optimization (IUO)}\label{alg:euclid}
	\begin{algorithmic}[1]
		\State \textbf{Initialization:} Set $d_{\rm{min}}=0$ and $d_{\rm{max}}=d^{ul}=\delta$ (large) such that $R^{ul}_{r, \mathit{\ell}} < C_\ell$ $\forall \ell$.
		\While{$\max_{\ell}(C_\ell - R^{ul}_{r, \mathit{\ell}}) > \epsilon ~\textbf{or}~ \max_{\ell}(\mathit{R^{ul}_{r, \mathit{\ell}}- C_\ell})>0$}
		\If{$\max_{\ell}(\mathit{R^{ul}_{r, \mathit{\ell}}- C_\ell})>0$}
		\State $d_{\rm{min}}=d^{ul}$
		\Else
		\State $d_{\rm{max}}=d^{ul}$
		\EndIf		  
		\State $d^{ul}=(d_{\rm{max}}+d_{\rm{min}})/2$
		\State $\mathbf{F}^{ul}_r=\rm{Chol}\mathit{(\frac{1}{d^{ul}} \mathbf{H}^{ul}\mathbf{P}^{ul}\mathbf{H}^{{ul}^\top}+\mathbf{I}_L(\frac{1}{d^{ul}}+\rm{1}))}$
		\State $\mathbf{A}_r^{ul}=\rm{LLL}(\mathit{\mathbf{F}^{ul}_r})$
		\State $R^{ul}_{r, \mathit{\ell}}=\frac{1}{2}\log^+({||\mathbf{F}^{ul}_r \mathbfit{a}_{r,l}^{ul}||}^2)$
		\EndWhile\label{euclidendwhile}
		\State $\mathbf{F}^{ul}_\psi=\text{Chol}({ {({\mathbf{P}^{ul}}^{-1}+\mathbf{H^{\mathit{ul^\top}}}{(\mathbf{I}_L+\mathbf{D}^{ul})}^{-1}\mathbf{H^{\mathit{ul}}})}^{-1}})$
		\State $\mathbf{A}_\psi^{ul}=\rm{LLL}(\mathit{\mathbf{F}^{ul}_\psi}\mathbf{W}^\top)$
		\State Calculate ${\sigma_j^{ul}}^2$ using \eqref{sigma_j_ul2} $\forall~ m \in\{1, \dots, M\}$.
		\State Calculate $\mathbfit{r}_{\psi}^{ul} = {[R_{\psi,1}^{ul}, \dots, R_{\psi,K}^{ul}]}^\top$ using \eqref{Rpsi_ul}.		
		\State \textbf{return} ($\mathbf{D}^{ul},\mathbf{A}_r^{ul}, \mathbf{A}_\psi^{ul},\mathbfit{r}_{\psi}^{ul}$)
	\end{algorithmic}
\end{algorithm}
\begin{algorithm}[t]
	\caption{Updating the Precoding Matrix $\mathbf{B}^{ul}$ Algorithm}\label{barriermerhod}
	\begin{algorithmic}[1]
		\State Set $\theta = \theta^{(0)} >0$, and solve the barrier problem using newton method to get $\mathbf{B}^{ul^{(0)}} = {\mathbf{B}^{ul}}^*(\theta)$. 
		\State For barrier parameter $\eta > 1$	and $\gamma = 1, 2, 3, \dots,$ repeat
		\State ~~~Solve barrier problem at $\theta = \theta^{(\gamma)}$ using newton method initialized at $\theta^{(\gamma-1)}$ to produce $\mathbf{B}^{ul^{(\gamma)}} = {\mathbf{B}^{ul}}^*(\theta)$.
		\State ~~~Stopping criteria at $K+L/\theta < \epsilon$
		\State ~~~Else, increase $\theta := \eta \theta$
		
		\State \textbf{return} ($\mathbf{B}^{ul}$)
	\end{algorithmic}
\end{algorithm}

\subsection{Downlink Rate Optimization}
The aim of this step is to obtain an achievable downlink user-rate greater or equal to the achievable one in the uplink, i.e., $\mathit{R}^{dl}_{\psi,\mathit{k}} \geq \mathit{R}^{ul}_{\psi,\mathit{k}}$. Given the downlink channel matrix $\mathbf{H}^{dl}=\mathbf{H}^{ul\top}$, the downlink powers $P^{dl}_{\ell}$, $\ell=1,\ldots,L$, and the outputs of the uplink optimization algorithm, the achievable downlink user-rate $\mathit{R}^{dl}_{\psi,\mathit{k}}$ should be tuned by selecting proper integer coefficient matrix $\mathbf{A}^{dl}_r$, beamforming matrix $\mathbf{B}^{dl}$, and downlink compression distortion covariance matrix $\mathbf{D}^{dl}$. One way to formulate this problem is as follows
\begin{equation}
\label{DL_opt}
\begin{split}
\mathop{\hbox{min}}_{\mathbf{A}^{dl}_r, \mathbf{B}^{dl}, {\mathbf{D}^{dl}}} & \sum\limits_{k=1}^{K}(R_{\psi,k}^{dl}- R_{\psi,k}^{ul})^2
 \cr {\hbox{subject to}}\quad & 
{p}^{dl}_\ell \leq P^{dl}_\ell, \text{ and } \mathit{R}^{dl}_{r,\mathit{\ell}}\leq C_\ell ~\forall~ \ell\cr 
\end{split}
\end{equation}
To simplify this problem we choose $\mathbf{A}^{dl}_r=\mathbf{A}^{ul\top}_r$. Then, we obtain $\mathbf{B}^{dl}$ that minimize \eqref{DL_opt} using the BFGS Quasi-Newton algorithm with a cubic line search procedure \cite{BFGS} and calculate the downlink distortion levels $\mathbfit{d}^{dl} ={[d_1^{dl},\dots, d_L^{dl}]}^\top$ using \eqref{comp_dl}. The line search in the BFGS algorithm must satisfy the Wolfe conditions in order to ensure sufficient step length taken in each search direction. Finally, the optimal $\mathbf{B}^{dl}$ matrix is obtained when the partial derivatives of $\mathbf{B}^{dl}$ are sufficiently too small \cite{BFGS}. If the constraints in \eqref{DL_opt} are not satisfied, we update the initial value of $\mathbf{B}^{dl}$, and  repeat until the constraints are satisfied. The details of this procedure are given in Algorithm~2.
\begin{algorithm}[t]
	\caption{Iterative downlink optimization (IDO)}\label{alg:euclid}
	\begin{algorithmic}[1]
		\State \textbf{Initialization:} set $\mathbf{H}^{dl}=\mathbf{H}^{ul^\top}$, $\mathbf{A}_r^{dl}=\mathbf{A}_r^{ul^\top}$,  $\mathbf{C}^{dl}=\mathbf{C}^{ul^\top}$, and $\mathbfit{r}_{\psi}^{dl} = {[R_{\psi,1}^{dl}, \dots, R_{\psi,K}^{dl}]}^\top=0_{K \times 1}$.
		\While{$\sum\limits_{k=1}^{K}|R_{\psi,k}^{dl}- R_{\psi,k}^{ul}| > \epsilon~\textbf{or}~ \max_{\ell}({p}^{dl}_\ell-P^{dl}_{\ell})>0 $}
		\State Set $\delta=0$, initialize $\mathbf{B}^{dl}_0 = L \times M$ matrix of i.i.d. $\mathcal{N}(0,1)$, and $\mathbf{\Gamma}_0=\mathbf{I}$.
		\While {termination condition for BFGS method} 
		\State Compute line search $\mathbf{\Theta}_\delta=-\mathbf{\Gamma}_\delta \nabla{f_{{un}}(\mathbf{B}^{dl}_\delta)}$, and step length $\gamma_{\delta} >0$.
		\State Calculate $\mathbf{B}^{dl}_{\delta+1} =\mathbf{B}^{dl}_\delta + \gamma_{\delta} \mathbf{\Theta}_\delta$ and $\mathbfit{d}^{dl}=\mathbf{C}^{dl} ~\boldsymbol{\xi}^{dl}$
		\State {Calculate ${p}_{\ell}^{dl}$, ${\rho}^{dl}_k$, and $\mathbfit{r}_{\psi}^{dl} = {[R_{\psi,1}^{dl}, \dots, R_{\psi,K}^{dl}]}^\top$ using \eqref{tilde_dl}, \eqref{rhok_dl}, and \eqref{Rpsi_dl}, respectively}
		\State Calculate $\boldsymbol{\beta}_\delta=\mathbf{B}^{dl}_{\delta+1}-\mathbf{B}^{dl}_{\delta}$. 
		\State Calculate $\boldsymbol{\Omega}_\delta = \nabla{	f_{un}(\mathbf{B}^{dl}_{\delta+1})}-\nabla{f_{un}(\mathbf{B}^{dl}_\delta)}$.	
		\State Update $\mathbf{\Gamma}_\delta$ using $\boldsymbol{\beta}_\delta$ and $\boldsymbol{\Omega}_\delta$ as in \cite{BFGS}.
		\State Set $\delta=\delta+1$. 
		\EndWhile 
		\EndWhile\label{euclidendwhile}	
		\State \textbf{return} ($ \mathbfit{r}_{\psi}^{ul}, \mathbfit{r}_{\psi}^{dl}$)
	\end{algorithmic}
\end{algorithm}
\section{Numerical Results}
\label{5}
In this section, the proposed MPTWR optimization scheme is evaluated and compared to the conventional  integer-forcing source and channel coding (IFSC+IFCC) scheme \cite{8635883} and the rate adaptive integer-forcing source and channel coding scheme (RAIFSC+IFCC) \cite{8362202} under the fronthaul capacity constraint per each RRH. We also compare with optimized Wyner-Ziv (WZ+IFCC) and single-user (SU+IFCC) compression schemes with integer-forcing channel coding. We use 5000 realizations of the $L \times K$ channel matrix $\mathbf{H} = \mathbf{H}^{ul} = \mathbf{H}^{dl^\top}$, where each element $h_{\ell,k}$ is i.i.d $\mathcal{N}(0,1)$. We set $L=2$ RRHs, $K=4$ users ($M=2$ user-pairs). It is assumed that the coding power of the coarse lattice $\Lambda_c$  is equal to $p^{ul} = \frac{10^{\rm{SNR}/10}}{K}$, where $\rm{SNR}$ is the signal-to-noise ratio in dB. Also, the capacities of the fronthaul links are assumed to be equal, i.e., $C_1 = C_2$. 

\begin{figure}[t]
	\centering 
	\includegraphics[width=9.3cm,height=4.5cm]{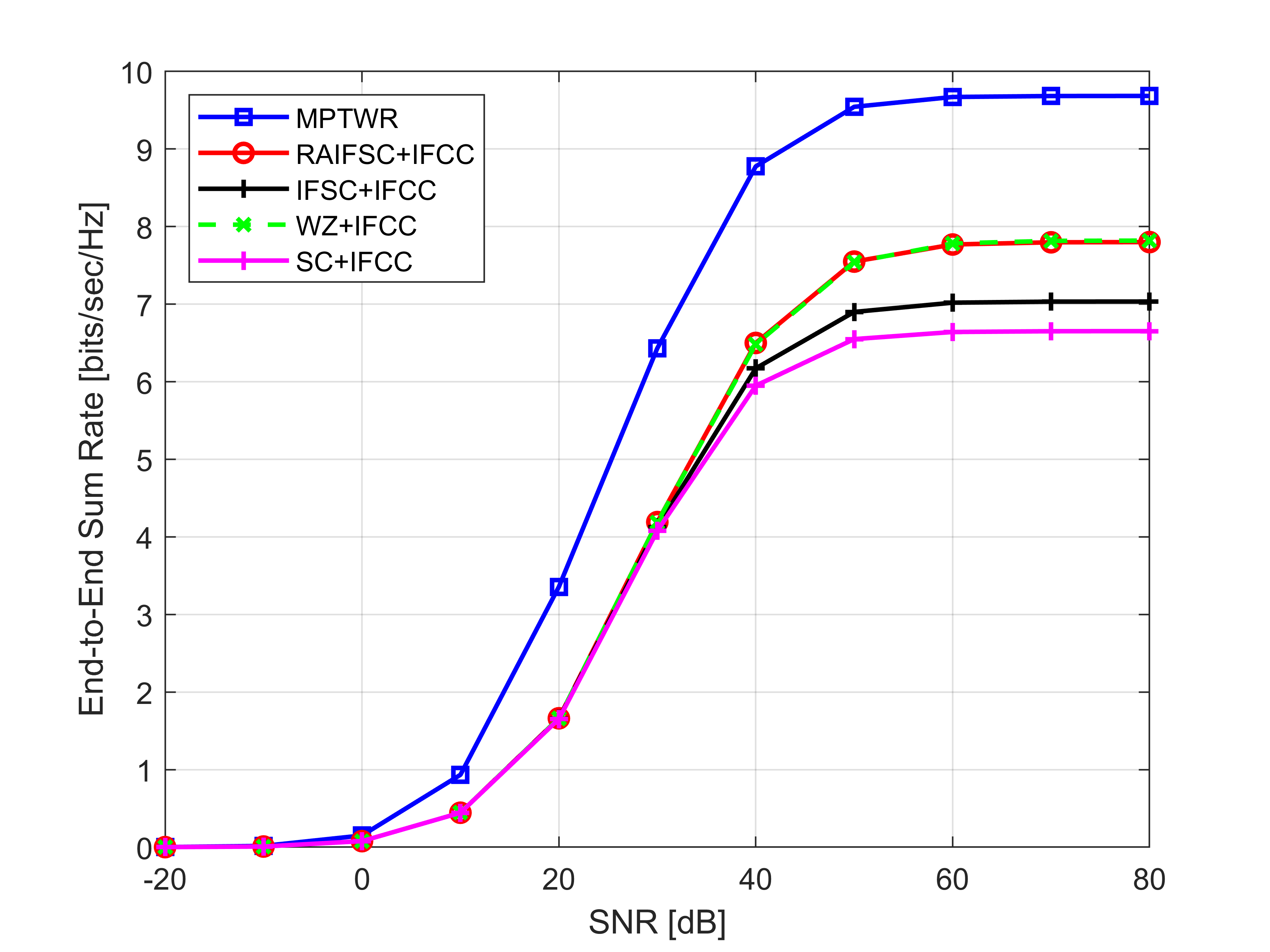}
	\caption{End-to-End sum rate versus SNR in dB at $C_\ell=4$ bits/transmission.}
	\label{Fig:RatevsSNR}
\end{figure}

\begin{figure}[t]
	\centering 
	\includegraphics[width=9.3cm,height=4.5cm]{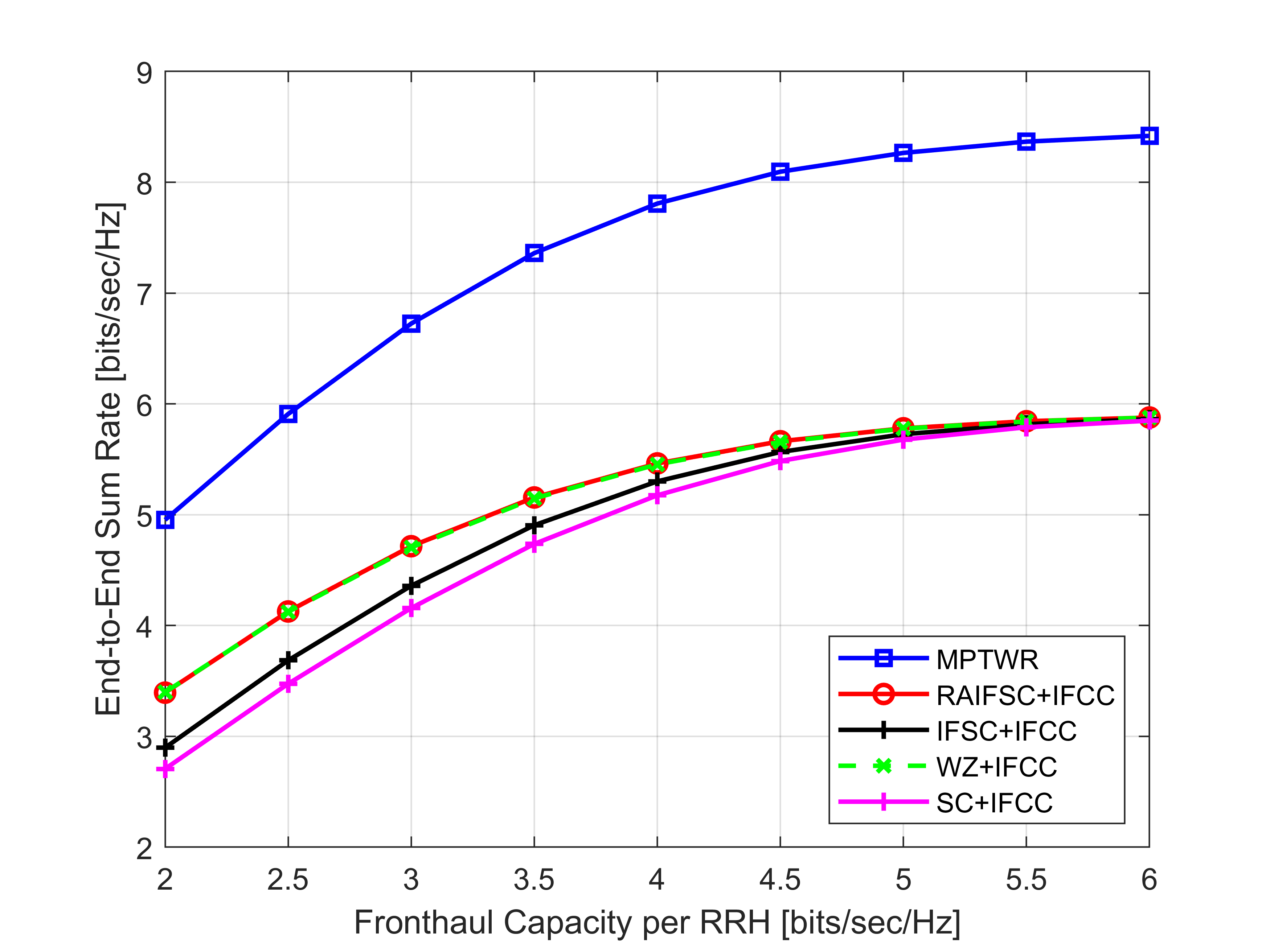}
	\caption{End-to-End sum rate versus $C_\ell$ in bits/transmission at SNR = 35 dB.}
	\label{Fig:RatevsCl}
\end{figure}

Fig. \ref{Fig:RatevsSNR} and \ref{Fig:RatevsCl} show the total achievable end-to-end rate of our proposed scheme and conventional ones in bits/sec/Hz versus SNR at  $C_\ell = 4$ bits/transmission and different fronthaul link capacity values at SNR of $35$ dB, respectively. These figures demonstrate that our proposed scheme has a superior performance over other conventional approaches. This is due to the exploitation of the multi-pair lattice-based computation strategy that reduces the number of decoded linear combinations at the BBU pool to 2 equations instead of the 4 equations required by other IF schemes. 
In addition, the performance of the RAIFSC+IFCC scheme is nearly the same as that in the optimized WZ scheme, while the performance of IFSC+IFCC scheme is close to them. Further, the optimized SU has the poorest performance as usual because
the correlation between the received signals at all RRHs is not
exploited.

\section {conclusion}
\label{6}
We proposed a multi-pair two-way user-rate optimization scheme for intra C-RAN communications, where users inside the network are grouped into communicating pairs. We used a multi-pair lattice-based computation strategy, where the BBU pool decodes integer linear combinations of paired users' codewords instead of decoding linear combinations of individual codewords. This reduces the required number of computation steps at the BBU pool, thereby reducing the number of rate constraints. In addition, instead of recovering the original messages as common in the BBU pool, the previously computed equations are compressed directly and forwarded to the RRHs through the fronthaul links. The scheme achieves significant improvement in the end-to-end rate compared to existing schemes.

\bibliographystyle{IEEEtran}
\bibliography{ref}

\end{document}